# JDLL: A library to run Deep Learning models on Java bioimage informatics platforms

Carlos García López de Haro[1, 2], Stéphane Dallongeville[1, 2], Thomas Musset[1, 2], Estibaliz Gómez de Mariscal[3], Daniel Sage[4], Wei Ouyang[5], Arrate Muñoz-Barrutia[6], Jean-Yves Tinevez[7, *], Jean-Christophe Olivo-Marin[1, 2, *]

[1]Bioimage Analysis Unit, Institut Pasteur, Université Paris Cité, Paris, France. [2]CNRS UMR 3691, Institut Pasteur, Paris, France. [3]Instituto Gulbenkian de Ciência, Lisbon, Portugal. [4]Biomedical Imaging Group and Center for Imaging, Ecole Polytechnique de Lausanne, Lausanne (EPFL), Switzerland. [5]Science for Life Laboratory, KTH Royal Institute of Technology, Stockholm, Sweden. [6]Biomedical Sciences and Engineering Laboratory, Universidad Carlos III de Madrid, Leganés, Spain. [7]Image Analysis Hub, Institut Pasteur, Université Paris Cité, Paris, France. Correspondence: [*]jean- yves.tinevez@pasteur; jean-christophe.olivo-marin@pasteur.fr

The advancements in Artificial Intelligence (AI) technology over the past decade have been a breakthrough on imaging for life sciences, paving the way for novel methods in image restoration [1], reconstruction [2], and segmentation [3]. However, the wide adoption of DL techniques by end-users in bioimage analysis is hindered by the complexity of their deployment. These techniques stem from a variety of rapidly evolving frameworks (e.g., TensorFlow 1 or 2, Pytorch, etc.) that come with distinct and often conflicting setups, which can discourage even proficient developers. Consequently, this has led to integration hassles or even absence in mainstream bioimage informatics platforms such as ImageJ, Icy, and Fiji, many of which are primarily developed in Java.

We present JDLL (Java Deep-Learning Library), a Java library that provides a comprehensive toolkit and application programming interface (API) for crafting advanced scientific applications and image analysis pipelines with DL capabilities. JDLL streamlines the installation, maintenance, and execution of DL models across any major DL frameworks. JDLL is based on two main components (Figure 1). The first, the DL engine installer, acts as a framework manager. It facilitates the downloading and integration of leading DL frameworks such as Tensorflow 1 & 2, PyTorch, and ONNX. It is capable of functioning with both CPU and GPU across a vast array of supported versions. Notably, the DL-engine-installer automatically identifies the required engine and version from the model specifications, handling downloads or loads seamlessly for the end user. The second component, DL-model-runner, offers a Java API for performing inference on models built on the aforementioned DL frameworks. Additionally, this component can connect to the Bioimage Model Zoo (BMZ) website [4], that shares pretrained models relevant for life sciences, facilitating their download or retrieval and running inference effortlessly. It can work around memory (CPU or GPU) limitations when processing large images, thanks to its tiling feature. Furthermore, it boasts a user-friendly API for model loading or downloading, presenting comprehensible details about its training process and the underlying DL framework.

The flexibility and adaptability of these components are pivotal in facilitating the effortless installation and utilization of the full spectrum of DL models prevalent in the life sciences. JDLL is streamlined, carrying minimal dependencies, namely ImgLib2 [5] and specific libraries for processing JSON and YAML files, commonly used for DL model specifications. It is built to be compatible with existing and potential future DL frameworks, leveraging ImgLib2 to create a framework-agnostic tensor implementation. Consequently, applications built on JDLL will not require engine-specific code. JDLL offers a universal, singular implementation compatible with all engines across their various versions. Its cutting-edge management of DL engines empowers it to operate multiple models within a single program, even if they were trained on normally incompatible DL engines. For example, JDLL can effortlessly sequence a style-transfer model operating on Tensorflow 2 with an instance segmentation model running on Tensorflow 1 within a single script.

The Bioimage Model Zoo (BMZ) [6] represents a collaborative endeavor aimed at streamlining the utilization and sharing of Deep Learning (DL) models in life sciences. It provides specifications for the DL models and the engines they require to run, along with pre- and post-processing steps. Several community partners in BMZ , such as Icy, ImageJ/Fiji (via deepImageJ), ZeroCostDL4Mic and ImJoy [7–10] already target end-user applications. JDLL aspires to serve as the common core component for DL across the Java-based platforms in this list, introducing distinctive features to support this integration. The BMZ specifications detail both pre- and post-processing steps, and for true reproducibility, it is imperative that these steps yield consistent outputs across all platforms. All the existing Java platforms rely on their own custom and implementation. However, capitalizing on the universal tensor capabilities provided by JDLL and ImagLib2, these processing steps are now addressed generically. As a result, platforms incorporating JDLL can depend on a shared codebase for all processes tied to running a DL model. This approach minimizes redundant coding efforts and ensures pixel-perfect reproducibility. As a first consumer of the JDLL, Icy stands out as the premiere software capable of interfacing with a diverse array of engines/frameworks: Tensorflow 1 and 2, PyTorch (starting from v1.4), and ONNX (beginning with v1.3). Fiji has also recently replaced its custom DL runner by JDLL via deepImageJ 3.0.

JDLL serves as a foundational library designed to support a growing community of developers in creating scientific tools for end-users (bioimage analysts, biological scientists). Furthermore, its efficient and streamlined API reduces the complexities of executing a DL model to just a handful of code lines. This makes it compatible for use in analysis scripts (such as those in MATLAB, Fiji, Icy) or through connectors that integrate it with visual programming languages (we can foresee KNIME and Icy Protocols). This also positions JDLL as an invaluable asset for the biologist community. Thanks to its generality and versatility, JDLL offers the means to fuel scientific Java image analysis platforms with Deep Learning capabilities. Given the popularity of these platforms within the biologist research community, JDLL stands poised to bolster the widespread adoption of Deep Learning in the realm of life sciences.

**Code availability**

The source code, documentation, tutorials and examples implementing JDLL can be found at https://github.com/bioimage-io/JDLL. JDLL is made available under the open-source Apache software license.

**Acknowledgements**

This work has been partially supported by the Agence Nationale de la Recherche through the LabEx IBEID (ANR-10-LABX-62-IBEID), the Institut Carnot Pasteur Microbes & Santé (ANR 16 CARN 0023-01), the programs PIA INCEPTION (ANR-16-CONV-0005) and France-BioImaging (ANR-10-INBS-04); by DIM ELICIT Région Ile-de-France; by the EC through the H2020-FET-OPEN-2018–2019-2020-01 Grant No 862840 ("FREE@POC"); and additional internal funding from the Bioimage Analysis unit and the Institut Pasteur; and by Ministerio de Ciencia, Innovación y Universidades, Agencia Estatal de Investigación, under Grant PID2019-109820RB-I00, MCIN / AEI / 10.13039/501100011033/, co-financed by European Regional Development Fund (ERDF), "A way of making Europe" and the European Commission through the Horizon Europe program (AI4LIFE project, grant agreement 101057970-AI4LIFE).

**Author contributions**

Code concept and design were done by C. G. L. H., J.-Y. T., S. D. and J.-C. O.-M. JDLL coding development and implementation was done by C. G. L. H., J.-Y. T., T. M. and S. D.; manuscript organizing and writing by C. G. L. H., J.-Y. T., A. M.-B. and J.-C. O.-M. with all authors contributing comments and revisions; funding and project administration by J.-Y. T. and J.-C. O.-M.

**Competing interests**

The authors declare no competing interests.

**Figure 1:** The JDLL architecture. JDLL is a Java API that can manage, and load models created with a wide range of Deep-Learning frameworks or engines (top). It can also download and deploy the models shared on the Bioimage Model Zoo (BMZ) repository following the BMZ conventions. These models are unpacked with JDLL, along with the engines required to run inference with the models, in a transparent manner to API consumers. JDLL provides to these consumers (such as Icy and deepImageJ, bottom) a simple and unified API to run inference on images via a custom ImgLib2 wrapper for their specific image data model. Because the API is generic and relies on an ImgLib2 component (see text), JDLL can be used by any Java software platform, fostering the reproducibility of Deep-Learning models' deployment.

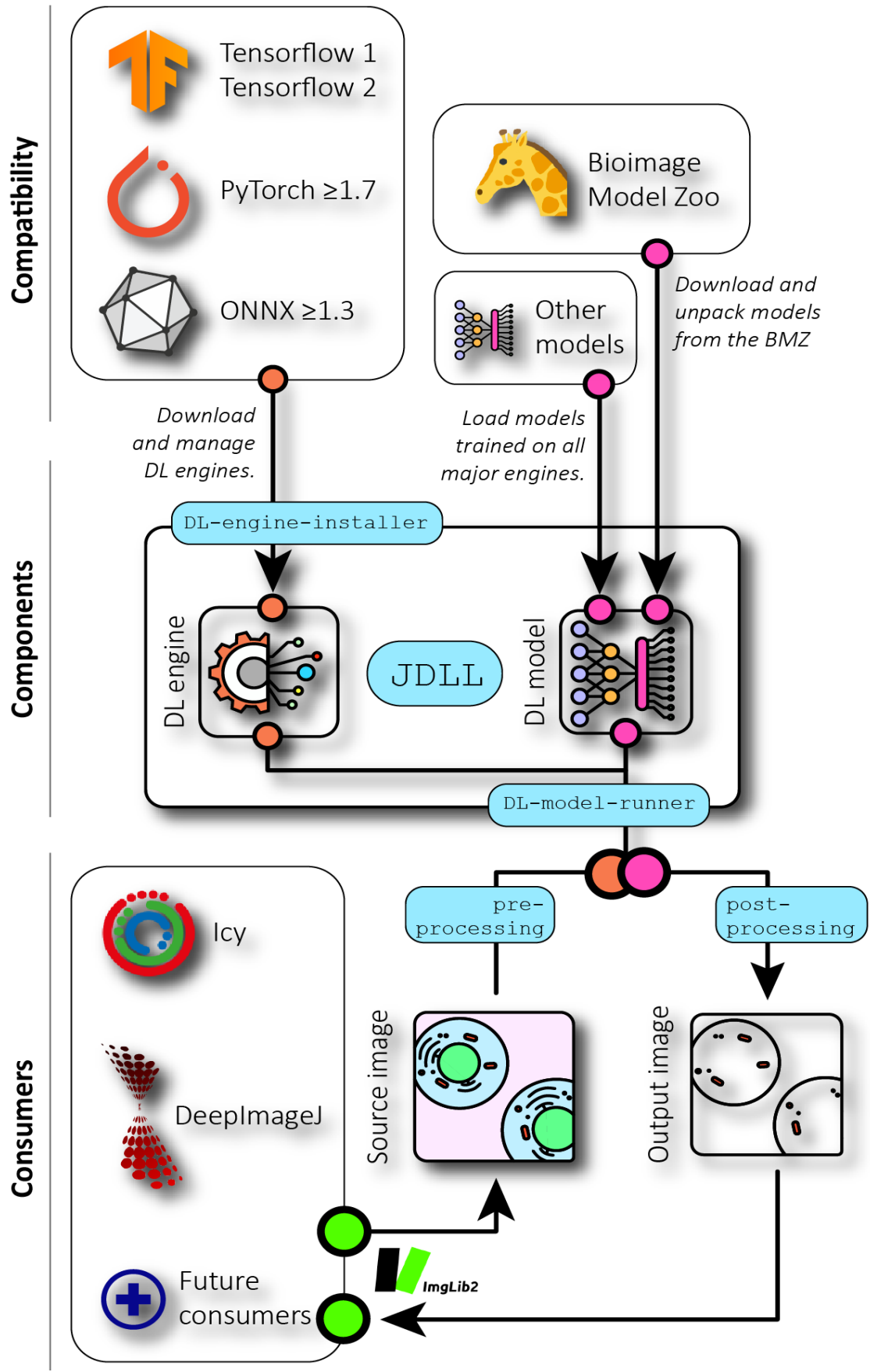

**Figure 1**